\documentstyle[art12]{article}
\begin{document}

\newcommand{\bean}{\begin{eqnarray*}}
\newcommand{\eean}{\end{eqnarray*}}
\newcommand{\ed}{\end{document}}
\newcommand{\pr}{\prime}
\newcommand{\ppr}{\prime\prime}
\newcommand{\cE}{{\tilde E}}
\newcommand{\vphi}{{\varphi}}
\newcommand{\oO}{O(k^{-1})}
\newcommand{\be}{\begin{equation}}
\newcommand{\ee}{\end{equation}}
\newcommand{\barr}{\begin{array}}
\newcommand{\earr}{\end{array}}
\newcommand{\bea}{\begin{eqnarray}}
\newcommand{\eea}{\end{eqnarray}}
\newcommand{\pa}{\partial}
\newcommand{\xx}{\hbox{}^*_*}
\newcommand{\sds}{\subset\hskip - 1em +}
\newcommand{\proof}{\medskip\noindent{\it Proof.}\quad }
\newcommand{\qed}{\hfill \fbox{}\medskip}

\title{New
realizations of
observables
in dynamical systems with second class constraints.}
\author{A.V.Bratchikov \\ Kuban
State Technological University,\\ 2 Moskovskaya Street, Krasnodar,
350072, Russia\\ E-mail:bratchikov@kubstu.ru} \date {} \maketitle

\begin{abstract}
In the Dirac bracket approach to dynamical
systems with second class constraints
observables are represented by elements of
a quotient Dirac bracket algebra. We describe
families of new realizations of
this algebra through quotients of the original Poisson  algebra.
Explicite expressions for generators and
brackets of the algebras under consideration are found.

\end{abstract} \bigskip



\newpage


\bigskip

\section{Introduction}
In a dynamical system with first class constraints
physical
functions
are elements of
a Poisson bracket algebra of first class functions
(see e.g. \cite{GT}).
Observables
are classes of the
physical functions
modulo the functions vanishing on constraint surface .

In the
Dirac bracket approach to a system  with second class
constraints \cite{D} the
original Poisson bracket is replaced by the Dirac one and constraints
become first class.In this case all the functions on phase space are
first class  and
observables
are elements of
the Dirac bracket algebra of  all the functions
modulo the functions vanishing on constraint surface.

     The latter quotient algebra can be realized as
a Poisson bracket algebra  of  the
functions on constraint surface
\cite {Sn}.
Another useful realization  can be obtained by using the Abelian
conversion
of second class constraints \cite {BT}.  The algebra  of  observables is
also realized as
a quotient of the original Poisson algebra of first class functions
\cite {Br}.

       The aim of this article is to present the new
Poisson algebras with respect to the original bracket
which are isomorphic to the algebra of observables in
a dynamical system with second class constraints.

The construction uses a family of nested subalgebras of the
original Poisson algebra of first class functions and their ideals
which are generated by the functions vanishing on constraint surface.
Existence of such subalgebras imposes some restrictions on
possible constraints. Solving the defining equations we find explicite
expressions for generators of the algebras under
consideration. This enables us to construct families of
new isomorphic images of
the algebra of observables.
The new algebras are Poisson ones with respect to the
original bracket.Using these realizations of a constrained
system one can avoid quantization of the Dirac bracket.

The paper is organized as follows.
In Section 2 we
review a description of a
system with second class constraints through the original Poisson
bracket. In Section 3
we find explicite expressions for the functions on phase space which
serve as generators of new Poisson algebras.  These algebras are
constructed and studied in Section 4.  In Section 5 we describe new
realizations of the algebra of observables in a system with second
class constraints.

\bigskip

\section {A  realization of
observables through the original Poisson bracket
}
Let $M$ be a phase space with the phase variables
$\eta_n,\,n=1...2N,$ and the Poisson bracket
$
[\eta_m,\eta_{n}
]=\omega_{mn}(\eta).
$ Let $H(\eta)$ be the original hamiltonian
and $\varphi_j(\eta), j=1...2J,$ the second class constraints
$det [\varphi_j,\varphi_{k} ]|_{\varphi=0} \ne 0.$
We shall assume \cite {D} that all the quantities vanishing on
constraint surface are linear functions of $(\varphi_{i}).$

Hamilton equations of the system under consideration are read
\bea
\label{hj}
\frac {d} {dt}
\eta_n = [\eta_n,H_T ],\qquad  \varphi_j=0.
\eea
Here
\bea
\label{H}
H_T=H+{\lambda_j\varphi_j}
\eea
and $\lambda_j=\lambda_j(\eta)$ are defined by
the  equations \bea \label  {hh12}
[H_T, \varphi_j]|_{\varphi=0}=0.
\eea

From (\ref {hj}) it follows
\bea
\label{hh1}
\frac {d} {dt}
f = [f,H_T ],\qquad  \varphi_j=0
\eea
for all $f=f(\eta).$

Using (\ref {hh12}) one can write equations
(\ref{hh1})
as
\bea \label{h3}
\frac {d} {dt}
f = [f,H_T ]_D,
\qquad \varphi_j=0.
\eea
Here the Dirac bracket was introduced
\bean \label{Dbr}
[g,h]_D=[g,h]- [g,\varphi_{j}]c_{jk}[\varphi_{k},h],\qquad
c_{jk}[\varphi_{k},\varphi_{l}]=\delta_{jl} .
\eean

The constraints $(\varphi_j)$ are first class with respect to
the Dirac bracket: $[\varphi_{j},\varphi_{k}]_D=0$  and
the physical functions are defined by the equations
\bean \label{hjkl}
{[f,\varphi_j]_D}|_{\varphi=0}=0
\eean
which are satisfied identically.
Let $A$ be the
space
of  functions on $M$
and $\Phi\subset A$ be the subspace of the functions
which vanish on constraint surface.Then the algebra of
observables is the Dirac bracket algebra  $A/\Phi.$ Note that
$A/\Phi$ is also an algebra with respect to the pointwise multiplication
and hence $A/\Phi$ is a Poisson algebra.

Let
$\{f\} \in
A /\Phi $ be the coset represented by $f \in A.$ Then, using
(\ref {h3}) and (\ref {hh12}) one can obtain the Hamilton equations for
observables \bea \label{h37} \frac {d} {dt} \{f\}=[\{f\},\{H\})]_D.
\eea

In a recent article \cite {Br} a new approach to
quantization of the system (\ref {hj})
was proposed.
Let $\Upsilon$ be the algebra of the functions which are
quadratic in $\varphi_j$ and let
$\Omega$ be the algebra of first class functions:
\bea
\label{hjkl}
\Omega =\{f\in A|\, [f,\varphi_j]|_{\varphi=0}=0\}.
\eea

For $u \in \Upsilon$ we have $u|_{\varphi=0}=0.$ The set $\Upsilon$
includes the element $\varphi_j\varphi_j$ and hence from the
equations $u=0,u\in \Upsilon,$ it follows $ \varphi_j=0.$
Thus, the constraints $(\varphi_j)$ and $\Upsilon$ are equivalent and
we can replace equations (\ref {hj}) by
\bea
\label{hhh1}
\frac {d} {dt}
f = [f,H_T ],\qquad
u_{ij}\varphi_i\varphi_j=0.
\eea
Here
$u_{ij}=u_{ij}(\eta)$ are arbitrary functions.
In contrast with the original ones the new constraints
$\Upsilon$ are first class.

In this approach the algebra of physical functions consists of all the
functions which satisfy the equations \bea \label{hjk} [f,u ]\in
\Upsilon \eea for all $u\in \Upsilon.$ One can show that these
equations are equivalent to the definition of first class functions
(\ref{hjkl})
in the original second class system.
Due to equations (\ref {hjk}) and (\ref {hjkl}) the algebra of
observables is the Poisson
algebra $\Omega/\Upsilon.$

Let
$\{f\}^\bullet \in \Omega/\Upsilon$ be the coset represented by $f \in
\Omega.$
Then the
equations for observables read
\bea \label{h777} \frac {d}
{dt} \{f\}^\bullet= [\{f\}^\bullet,\{H_T\}^\bullet].  \eea

The present approach and the Dirac bracket one are
related by the isomorphism of the algebras of observables
$\Omega/\Upsilon$ and
$A/\Phi$ \cite {Br}:
\bean
T(\{g\}^\bullet)= \{g\}.
\eean

Below we shall obtain new realizations of the algebra
$A/\Phi$ through quotients of the original Poisson algebra.

\bigskip
\section {Generators of $\Omega_{s+1}$}
Let   $\Omega_{s+1},\, s\in N,$ be the space of the functions on $M$
which are defined by the equations
\bea \label{Un} [\varphi_j,\tilde g]\in  \Upsilon_{s}.
\eea
Here
\bean \Upsilon_{s}=\{u\in A |u=u_{j_1\ldots j_{s}}
\varphi_{j_1}\ldots \varphi_{j_{s}},u_{j_1\ldots j_{s}}(\eta)\in A\}.
\eean
It is seen that
$\Upsilon_{s+1}\subset \Omega_{s+1} \subset \Omega_{s},$ $\Omega_2=\Omega,
\Upsilon_{s+1}\subset \Upsilon_{s},
\Upsilon_{2}=\Upsilon$ and
$\Upsilon_{1}=\Phi.$
We shall denote $\Upsilon_{0}=\Omega_1=A.$

To describe elements of  $\Omega_{s+1}$ explicitly let us consider
equations (\ref {Un}) with the initial condition
\bea \label {ic1}  \tilde g{(\eta)} \in \{g(\eta)\}
\eea
for some $g\in A.$
A solution to these equations
can be represented in the form
\bea  \label{geg}  \tilde
g
=g
+ \sum_{r=1}^s { \frac 1 {r!}
\nu_{i_1\ldots i_r}(\eta)
\varphi_{i_1}
\ldots \varphi_{i_r}}+
\nu_{i_1\ldots i_{s+1}}(\eta)
\varphi_{i_1}
\ldots \varphi_{i_{s+1}}.
\eea
Note that the last term of (\ref {geg}) satisfies (\ref {Un})
for arbitrary $\nu_{i_1\ldots i_{s+1}}.$

We shall assume that  $\nu_{i_1\ldots i_r}
,r=1\ldots s,$ is
symmetric:
\bea  \label{gegg}
\nu_{\underbrace{...i_a...i_b...}_r}-
\nu_{\underbrace{...i_b...i_a...}_r}
\in  \Upsilon_{p+1-r},\quad p \ge s.
\eea

Substituting (\ref {geg}) into  (\ref {Un}) and using (\ref {gegg}) we
get \bean \label{ge} [\varphi_{j},g]+ \nu_{i_1}
[\varphi_{j},\varphi_{i_1}]+
\sum_{r=2}^{s} {\frac 1
{(r-1)!}\left(
[\varphi_{j},\nu_{i_1\ldots i_{r-1}}]+
\nu_{i_1\ldots
i_{r}}[\varphi_{j},\varphi_{i_{r}}]
\right)\varphi_{i_1}\ldots \varphi_{i_{r-1}}} \in  \Upsilon_{s}.
\eean
It is easy to see that a solution to these equations is
\bea
\label{rrrr}
\nu_{i_1\ldots i_r}=
(-1)^r D_{i_r}...D_{i_1}g,
\qquad r=1\ldots s.
\eea
Here $D_{i}=c_{ij}[\varphi_{j},\cdot].$

One can check that $D_{i}$ satisfy the commutator relations
\bea
\label {comrel}
D_{i}D_{j}-D_{j}D_{i}=[c_{ij},\cdot]_D
\eea
and for $u\in \Upsilon_{r}$
\bea
\label {comrels}
D_{i}u \in  \Upsilon_{r-1}.
\eea

Now let us consider equations (\ref {gegg}).
It is sufficient to find a solution to  these equations for
$a=k+1,b=k,k=1\ldots r-1.$ Substituting (\ref {rrrr}) into (\ref
{gegg}) and using (\ref {comrel})  we have \bea \label{rrrr5}
D_{i_{r}}\ldots D_{i_{k+2}}[c_{i_{k+1}i_k},D_{i_{k-1}}\ldots
D_{i_{1}}g]_D \in \Upsilon_{p+1-r}, \qquad k=1\ldots r-1.  \eea A
solution to these equations is given by \bea \label {cw} c_{ij}=
\psi_{ij}(\varphi)+v_{ij}, \qquad
v_{ij}(\eta)\in \Upsilon_{p-1}.
\eea
Here $\psi_{ij}$ is a function of the constraints $(\varphi_{j})$
only.

To check that  $c_{ij}$ satisfy equations (\ref {rrrr5})
observe that for $f\in A$
\bean
\label {ir}
[c_{ij},f]_D\in
\Upsilon_{p-1} \eean
and due to (\ref {comrels})
\bean
\label{rrrr55}
D_{i_{r}}\ldots D_{i_{k+2}}[c_{i_{k+1}i_k},D_{i_{k-1}}\ldots
D_{i_{1}}g]_D\in \Upsilon_{p+k-r} \subset  \Upsilon_{p+1-r}
\eean
for all $k=1\ldots r-1.$
Thus for $c_{ij}$ (\ref {cw}) expressions (\ref {geg}, \ref
{rrrr}) give us  a solution to equations (\ref {Un}) with the initial
condition (\ref {ic1}).

      Let now $\tilde g'$ be another solution  to equations
(\ref {Un}) with the same initial condition $\tilde g'\in \{g\}.$
Then
$\sigma = \tilde g
- \tilde g'$ is a solution to (\ref {Un})
\bea
\label{Un22} [\varphi_j,\sigma]\in  \Upsilon_{s}
\eea
and
$\sigma =
\sigma_i\varphi_i$ for some $\sigma_i= \sigma_i(\eta).$

From (\ref{Un22}) it follows
\bea
\label{Un23}
[\varphi_{j_1},\ldots ,[\varphi_{j_{m-1}},
[\varphi_j,\sigma]]]\in \Upsilon_{s-m+1}.
\eea
Assume that
\bea \label{s1}
\sigma =
\sigma_{i_1\ldots i_m}(\eta) \varphi_{i_1}\ldots \varphi_{i_m}.
\eea
Substituting (\ref {s1}) into (\ref {Un23}) for $m\le s$ we
get $\sigma_{i_1\ldots i_m}|_ {\varphi=0}=0$ and hence  $ \sigma =
\sigma_{i_1\ldots i_{m+1}}(\eta) \varphi_{i_1}\ldots
\varphi_{i_{m+1}}.$

For $m=s$ \bea \sigma = \sigma_{i_1\ldots
i_{s+1}}(\eta) \varphi_{i_1}\ldots \varphi_{i_{s+1}}.  \eea

We have proved the proposition:

\vspace{3mm}
\noindent
PROPOSITION 3.1.
{\it For $c_{ij}$ (\ref {cw}) and $g \in A$ the set
$ \{g\}\cap \Omega_{s+1},\,s=1\ldots p,$ consists of all the
expressions \bea \label{geg3}  \tilde g =g + \sum_{r=1}^s { \frac
{(-1)^r} {r!} \left(D_{i_r}...D_{i_1}g\right) \varphi_{i_1} \ldots
\varphi_{i_r}}+ \nu_{i_1\ldots i_{s+1}} \varphi_{i_1} \ldots
\varphi_{i_{s+1}},
\eea
where $\nu_{i_1\ldots i_{s+1}}(\eta)$ are arbitrary functions.}

\vspace{3mm}
\noindent
In what follows we shall assume  that $c_{ij}$ is given by (\ref {cw})
and $1 \le s \le p.$

It is convenient to introduce the notation
\bean
L_s(g)=
g+
\sum_{r=1}^s {\frac {(-1)^r} {r!}
\left(D_{i_{r}}...D_{i_{1}}g\right)\varphi_{i_1}\ldots
\varphi_{i_r}}.
\eean.

The hamiltonian in
$\Omega_{s+1}$ is
\bea \label {ham} \tilde H_T=
L_s(H)
+u,\qquad u\in
\Upsilon_{s+1}
\eea
It can be represented in the form (\ref {H}),satisfies equation
(\ref {hh12}) and hence belongs to the family of admissible
hamiltonians.

\section {Algebraic properties of $\Omega_{s+1}$}
\vspace{3mm}
\noindent
PROPOSITION 4.1.
{\it $\Omega_{s+1}$ is an algebra
and $\Upsilon_{s+1}$ is an ideal of $\Omega_{s+1}$
with respect to
the original Poisson bracket ,
Dirac bracket and  pointwise multiplication.}


The proof is straightforward.

\vspace{3mm}
\noindent

Due to this proposition $\Omega_{s+1},\Upsilon_{s+1}$ and
$\Omega_{s+1}/\Upsilon_{s+1}$
are Poisson algebras with respect to $[\cdot,\cdot]_D$ as well as
$[\cdot,\cdot].$

Let
\bea  \label {fcf12}
\tilde g_a=
L_s(g_a) + u_a,\qquad  u_a \in \Upsilon_{s+1},
\eea
$a=1,2,$ be some elements of $\Omega_{s+1}$ and let $
\{\tilde g_a\}_{s} \in
\Omega_{s+1}/\Upsilon_{s+1}$
be the coset represented by $\tilde g_a \in \Omega_{s+1}.$

\vspace{3mm}
\noindent
PROPOSITION 4.2.
{\it For $\tilde g_1,\tilde g_2$
} (\ref {fcf12})
{\it one has}
\bea \label {cr}
[\tilde g_1,\tilde g_2]=L_s([g_1,g_2]_D)+\tilde u_{12},
\quad
[\tilde g_1,\tilde g_2]_D=L_s([g_1,g_2]_D)+\tilde v_{12},
\eea
\bean \label {cr4}
\tilde g_1\tilde g_2=L_s(g_1g_2)+\tilde w_{12},\qquad
\tilde u_{12},\tilde v_{12},\tilde w_{12},\in \Upsilon_{s+1}.
\eean

\proof
One can check that $[\tilde g_1,\tilde
g_2]$ satisfies equations (\ref {Un}) with the initial condition
$[\tilde g_1,\tilde g_2]\in  \{[g_1,g_2]_D\}.
$
Due to results of the previous section
one has
\bean \label {cr5}
[\tilde g_1,\tilde g_2]=L_s([g_1,g_2]_D)+\tilde u_{12},\qquad
\tilde u_{12}\in \Upsilon_{s+1}.  \eean

Other statements of the proposition are proved by using similar
arguments.

\qed

\vspace{3mm}
\noindent

COROLLARY 4.3.
{\it The Dirac bracket algebra $\Omega_{s+1}/\Upsilon_{s+1}$ is
isomorphic to the
 algebra  $\Omega_{s+1}/\Upsilon_{s+1}$
with respect to the original Poisson bracket.}

\proof
From equations
(\ref {cr}) we have
\bean
\label {guu}
[\{\tilde g_1\}_{s},\{\tilde
g_2\}_{s}]=
[\{\tilde g_1\}_{s},\{\tilde
g_2\}_{s}]_D=\{L_s([g_1,g_2]_D)\}_{s}.  \eean
\qed

\section {New realizations of observables }

\vspace{3mm}
\noindent
THEOREM 5.1.

(i){\it The Dirac bracket algebra $A /\Phi$ is isomorphic to
the algebra $\Omega_{s+1} /\Upsilon_{s+1}$ with respect to the
original Poisson
bracket.}

(ii){\it $A /\Phi$ and $\Omega_{s+1} /\Upsilon_{s+1}$ are
isomorphic with respect to the pointwise multiplication.}

\proof

Let us define the
linear function $T_{s}:  \Omega_{s+1} /\Upsilon_{s+1}\to A /\Phi$ \bean
\label {T} T_{s}(\{g\}_{s})=\{g\} .\eean
Each function  $g'\in \{g\}\bigcap \Omega_{s+1}$ can be written in the
form (\ref {geg3}).
Hence the inverse function $T^{-1}_s: A /\Phi \to
\Omega_{s+1} /\Upsilon_{s+1}$ is given by
\bean \label {T-1}
T^{-1}_s(\{g\})=\{L_s(g)\}_{s}
 .\eean
Computations
show that $T_s$ is the homomorphism
\bean
\label {mom23} T_s([\{g\}_s,\{f\}_s])= [T_s(\{g\}_s),T_s(\{f\}_s)]_D
\eean
and hence
$A /\Phi$ and  $\Omega_{s+1} /\Upsilon_{s+1}$
are isomorphic.

To prove the second statement we observe that $T_s$
is the homomorphism with respect to the pointwise multiplication:
\bean \label
{mom3} T_s(\{g\}_s\{f\}_s)=
T_s(\{gf\}_s)=\{gf\}=\{g\}\{f\}=T_s(\{g\}_s)T_s(\{f\}_s).
\eean
\qed

\vspace{3mm}
\noindent
COROLLARY 5.1.
{\it  $\Omega_{s+1} /\Upsilon_{s+1},s=1\ldots
p,$ are isomorphic to each other as Poisson algebras .}

\vspace{3mm}
\noindent
The function $T_{s+k,s}
$ which defines isomorphism
between $\Omega_{s+k+1} /\Upsilon_{s+k+1},k\ge 0,$ and $\Omega_{s+1}
/\Upsilon_{s+1}$
is given by \bean
T_{s+k,s}(\{g\}_{s+k})=\{g\}_{s}.
\eean

\vspace{3mm}
Theorem 5.1. gives us new realizations of the algebra of
observables $A /\Phi$ through the original Poisson bracket.
For a given system we have $p$ realizations,where $p$ is defined by
the form of $c_{ij}$ (\ref {cw}).

For $p=1$ the matrix $c_{ij}$ is arbitrary and there is only
one realization $\Omega_2 /\Upsilon_2=\Omega /\Upsilon.$
For $p=2$
\bean
\label {cww}
c_{ij}= \psi_{ij}(\varphi)+v_{ijk}
\varphi_k,
\qquad
v_{ijk}=v_{ijk}(\eta).
\eean
In this case the observables can be realized by
$\Omega /\Upsilon$ or $\Omega_3 /\Upsilon_3.$
When $c_{ij}=\psi_{ij}(\varphi)$ there
is an  infinite series of such realizations. Number $p$ can be used
for classification of second class constraints.

According to (\ref {hh1}) and (\ref {hh12}) the  Hamilton
equation in $\Omega_{s+1} /\Upsilon_{s+1}$ is \bean
\label{h77} \frac {d} {dt} \{{f}\}_s= [\{f\}_s, \{\tilde H_T\}_s].
\eean
Here $\tilde H_T$ is given by (\ref {ham}).

\section {Conclusion}
In the present article we have obtained new realizations of
observables in dynamical systems with second class constraints.
The observables are realized as Poisson algebras with respect to
the original bracket.
We have found the restrictions which are
imposed on constraints by consruction of such algebras.
The number of possible realizations of the observables for a given
system can
be used for classification of  second class constraints.
We have obtained explicite expressions for generators and brackets of
all the algebras under consideration.

\bigskip

{\bf Acknowledgements}

Author thanks I.V.Tyutin for reading the manusctipt and helpful
comments. The research was suppoted in part by RFBR grant 03-02-96521.


\bigskip

\begin{thebibliography}{99}
\bibitem {GT} Gitman, D.M. and Tyutin, I.V.:{\it  Canonical Quantization
of Fields with Constraints}, Nauka, Moscow,1986 (in Russian),
Springer Series in Nuclear Particle Phys.,Springer-Verlag, Berlin,1990.
\bibitem {D}P.A.M.Dirac,
{\it Lectures on Quantum Mechanics} (Yeshiva University, New York,1964).
\bibitem {Sn}I.J.Sniatycki,{\it Ann.Inst.H.Poincare
} {\bf 20}(1974) 365-372.
\bibitem{BT} I.Batalin and I.Tyutin,{\it Int.J.Mod.Phys.}
{\bf A6} (1991) 3255-3282.
\bibitem{Br} A.V.Bratchikov,{\it Lett.Math.Phys.}
{\bf 61} (2002) 107-111.
\end{thebibliography}
\end{document}